\documentclass[prd,aps,twocolumn,superscriptaddress,preprintnumbers,nofootinbib,showpacs]{revtex4}
\usepackage{graphicx}
\usepackage{exscale}
\usepackage[intlimits]{amsmath}
\usepackage{amsfonts}
\usepackage{amssymb,amscd}
\usepackage{epsfig}

\newcommand{\beqa}{\begin{eqnarray}}
\newcommand{\eeqa}{\end{eqnarray}}
\newcommand{\beq}{\begin{equation}}
\newcommand{\eeq}{\end{equation}}

\newcommand{\p}{\partial}
\newcommand{\too}{\rightarrow}

\begin{document}
\preprint{\parbox[t]{\textwidth}
{\small NYU/08/06 \hspace*{5mm} TP/08/20 \hfill hep-ph/yymmnnn}}

\title{Equation of State of Gluon Plasma from Local Action}

%\author{Christian~S.~Fischer}WW
%\affiliation{IPPP, University of Durham, Durham DH1 3LE, U.K.}

\author{Daniel~Zwanziger}
\affiliation{New York University, New York, NY 10003}
\date{Oct. 9, 2007}
\begin{abstract}
We review recent analytic and numerical results concerning the confinement scenario in Coulomb gauge.  We then consider a local, renormalizable, BRST-invariant action for QCD in Coulomb gauge that contains auxiliary bose and fermi ghost fields and sources.  When the auxiliary fields are integrated out, one obtains the standard Coulomb gauge action with a cut-off at the Gribov horizon.  We use the local formulation to calculate the leading correction to the Stefan-Boltzmann equation of state at high temperature due to the cut-off at the Gribov horizon.  It is of order~$g^6$, which is precisely the order at which the infrared divergence found by Lind\'{e} divergence first occurs.  No such divergence arises in the present calculation because the propagator of would-be physical gluons is suppressed in the infrared due to the proximity of the Gribov horizon in infrared directions.
\end{abstract}


\maketitle

\section{Introduction}

\subsection{Confinement scenario in Coulomb gauge}

	One would like to explain the presence of a long-range force that confines colored objects.  At the same time one is faced by the apparently contradictory requirement that the massless gluons that are supposed to transmit this force are absent from the physical spectrum.  
	
	This paradox is addressed in a scenario in Coulomb gauge that was originally developed by Gribov~\cite{Gribov:1977wm}.  In this scenario, the 3-dimensionally transverse, would-be physical gluon propagator, $D_{A_iA_j}({\bf k}, k_0)$, is suppressed at small $\bf k$ because of the proximity of the Gribov horizon in infrared directions, so physical gluons are absent from the physical spectrum.  On the other hand confinement of quarks, or any colored object, is explained by the long range of the instantaneous part, $V_{\rm coul}(R)$, of the time-time component of the gluon propagator, 
\beq
\label{defVcoul}
g^2 D_{A_0 A_0}({\bf x}, t) = V_{\rm coul}(|{\bf x}|)\delta(t) + P({\bf x}, t),
\eeq 
that couples universally to color charge.  Here $P({\bf x}, t)$ is a non-instantaneous vacuum polarization term that is screening, whereas the color-Coulomb potential, $V_{\rm coul}(R)$, is anti-screening \cite{Cucchieri:2001b}.  The concentration of probability at the Gribov horizon causes $V_{\rm coul}(R)$ to have the long-range, confining property $\lim_{R \too \infty} V_{\rm coul}(R) = \infty$.  The  different behavior of space and time components of the gluon propagator is possible because manifest Lorentz invariance does not hold in Coulomb gauge.  

	The local, renormalizable theory which will be presented here yields results in agreement with this scenario.  For example, at tree level, the 3-dimensionally transverse gluon propagator,
\beq
\label{transprop}
D({\bf k}, k_0) = 
{ {\bf k}^2 \over (k_0^2 + {\bf k}^2) {\bf k}^2 + m^4},
\eeq	 
is strongly suppressed in the infrared, vanishing like ${\bf k}^2$.  It has poles at
\beq
\label{poles}
- k_0^2 = E^2({\bf k}) \equiv {\bf k}^2 + {m^4 \over {\bf k}^2}, 
\eeq  
in agreement with the Coulomb-gauge energy obtained by Gribov. 

	All states are physical in Coulomb gauge, and the instantaneous part, $V_{\rm coul}(R)$, of the gluon propagator is directly related to the energy of a quark pair state, $|\Psi_{\bar q q}\rangle =  \bar q(0) q(R) |\Psi_0\rangle$.  This state in Coulomb gauge contains massive quark-anti-quark sources at separation $R$, and $\Psi_0$ is the vacuum state of pure glue.  $\Psi_{\bar q q}$ is invariant under the remnant gauge symmetry of time-dependent, space-independent gauge transformations $g(t)$ that is left unfixed by the Coulomb gauge condition,
\beq
\label{gaugecond}	
\sum_{i=1}^3 \p_i A_i({\bf x}, t) = 0.
\eeq
Let $L({\bf x}, T)$ be the time-like Wilson (not Polyakov) line\footnote{Here we follow the discussion of \cite{Greensite:2004a}.}
\beq
L({\bf x}, T) \equiv P 
\exp\Big[ \int_0^T dx_0 \ g t^bA_0^b({\bf x} , x_0)\Big],
\eeq
for quarks in an irreducible representation $r$ of SU(N), of dimension $d_r$.  Here $t^a$ is a basis of the Lie algebra of SU(N), $[ t^a, t^b] = f^{abc} t^c$ in the representation $r$.  The correlator of two Wilson lines, 
\beq
G(R,T) \equiv d_r^{-1} 
\langle \ {\rm Tr} [ L^\dag ({\bf x}, T) L({\bf y}, T)],
\eeq
may be expressed in terms of the Hamiltonian $H$ and the quark-pair state,
\beq
G(R,T) = 
\langle \Psi_{\bar q q}|\ 
e^{- (H - E_0)T} \ 
 |\Psi_{\bar q q}\rangle,
\eeq	
where $E_0$ is the vacuum energy, and $R = |{\bf x - y}|$.  We have
\beq
G(R, T) = \sum_n c_n e^{- \epsilon_nT},
\eeq
where $\epsilon_n = E_n - E_0$ is an excitation energy and $c_n = | \langle \Psi_n | \Psi_{\bar q q} \rangle |^2$ is a positive coefficient.  The logarithmic derivative,
\beqa
\label{logder}
\bar\epsilon(R, T) \equiv - { \p \ln[G(R, T)] \over \p T} 
\nonumber  \\
= { \sum_n \epsilon_n \ c_n e^{- \epsilon_nT} \over \sum_n c_n e^{- \epsilon_nT} },
\eeqa
yields the mean excitation energy above the vacuum energy of the quark-pair state $e^{-(H-E_0)T/2} \Psi_{\bar q q}$.  

	For small~$T$, the Wilson line is given by
\beq
L({\bf x}, T) = 1 + g \int_0^T dx_0 t^b A_0^b({\bf x}, x_0) + ... 
\eeq
and 
\beqa
G(R, T) & = & 1 - T E_{se} + (g^2/d_r) {\rm Tr}(t^b b^c)
   \\     \nonumber 
&& \times  \int_0^T \int_0^T dx_0 dy_0 \langle A_0^b({\bf x}, x_0) 
A_0^c({\bf y}, y_0) \rangle + ... \ , 
\eeqa
where $E_{se}$ is the regularized self-energy of the two quarks.  It might be thought that the integral is of order $T^2$.  However, with
\beq
\label{defprop}
\langle A_0^b({\bf x}, x_0) 
A_0^c({\bf y}, y_0) \rangle = \delta^{bc} D_{A_0 A_0}({\bf x - y}, x_0 - y_0),
\eeq
the instantaneous term in (\ref{defVcoul}), proportional to $\delta(x_0 - y_0)$, gives a contribution of order $T$, and we obtain, for small~$T$
\beq
G(R, T) = 1 - T [E_{se} - C_r V_{\rm coul}(R)] + ... \ ,
\eeq
where $C_r = - t^b t^b > 0 $ is the Casimir invariant in the representation $r$ of SU(N).  This gives, 
\beq
\label{zeroT}
\lim_{T \to 0} \bar\epsilon(R, T) = E_{se} - C_r V_{\rm coul}(R). 
\eeq 
[The minus sign in front of $V_{\rm coul}(R)$ occurs because, by~(\ref{defVcoul}) and~(\ref{defprop}), $V_{\rm coul}(R)$ is the color-Coulomb potential between {\it like} charges.]  Thus the {\em physical} energy of the quark-pair state $\Psi_{\bar q q}$ is given by the instantaneous part of the gluon propagator in Coulomb gauge.  Consistent with this, 
the color-Coulomb potential $V_{\rm coul}(R)$ and, more generally, $gA_0$ are renormalization-group invariants~\cite{Zwanziger:1998ez}.

	It may be shown that the remnant gauge invariance of time-dependent gauge transformations $g(t)$ is spontaneously broken, or not, according as $V_{\rm coul}(R)$ is not, or is, confining \cite{Greensite:2004a}.
	
	With these results, it is simple to establish that there is ``no confinement without Coulomb confinement" \cite{Zwanziger:2002sh}.  One sees from (\ref{logder}) that, for large T, $\bar\epsilon(R, T)$ converges to the minimum energy of a quark pair in the representation~$r$
\beq
\label{infiniteT}
\lim_{T \to \infty} \bar\epsilon(R, T) = V_r(R) + E_{se},
\eeq
which, for a confining representation,  is the energy of the flux-tube ground state plus $E_{se}$.  One also sees from~(\ref{logder}) that the time derivative of $\bar\epsilon(R, T)$ gives the negative of the variance of the excitation energy
\beq	
 { \p \bar\epsilon(R, T) \over \p T} = - ( \overline{\epsilon^2} 
- \bar\epsilon^2 ) < 0,
\eeq  
which is negative.  Thus $\bar\epsilon(R, T)$ is a monotonically decreasing function of $T$, and so $\bar\epsilon(R, 0) > \bar\epsilon(R, \infty)$.  We conclude from (\ref{zeroT}) and (\ref{infiniteT}) that if 
the quark-pair energy $V_r(R)$ in some representation $r$ is confining, $\lim_{R \to \infty}V_r(R) \to \infty$, then
\beq
\label{boundV}
 V_r(R) < - C_r V_{\rm coul}(R)   \ \ \ \ \ \ \ \   R \to \infty.
\eeq
Thus if the quark-pair energy is confining for some representation $r$, then also the color-Coulomb potential is confining $ - V_{\rm coul}(R) \to \infty$.  This analytic result strongly supports the confinement scenario in Coulomb gauge.

\subsection{Recent numerical results}
	
	Numerical studies provide a valuable laboratory for testing various confinement scenarios.  In accord with the above scenario, it was found \cite{CucchieriZ2000} that the equal-time would-be physical gluon propagator $D_{A_i A_j}({\bf k})|_{t=0}$ is indeed suppressed at small $\bf k$, while the fourier transform $\widetilde V_{\rm coul}({\bf k})$ is enhanced at small $\bf k$, which corresponds to a long-range, color-Coulomb potential $V_{\rm coul}(R)$. It was subsequently found numerically \cite{Greensite:2003xf} that the long-distance behavior of the color-Coulomb potential is consistent with  a linear increase at large R,
\beq
\label{linearcoul}
V_{\rm coul}(R) \sim \sigma_{\rm coul} R,
\eeq
with the Coulomb string tension given approximately by $\sigma_{\rm coul} \sim 3 \sigma$.  Here~$\sigma$ is the physical string tension between a pair of external quarks as determined from a large Wilson loop in the fundamental representation.   Other studies provided additional support for the confinement scenario in Coulomb gauge, including in particular the consistency of Coulomb-gauge and center-vortex scenarios \cite{Greensite:2004a, Greensite:2004b}.  These numerical studies were certainly encouraging for the confinement scenario in Coulomb gauge.  
	
	However, as in the old tale of the sorcerer's apprentice, it was unexpectedly discovered that this ``confinement scenario" works also in the {\em deconfined} phase.  Indeed it was found numerically that the long-distance behavior of $V_{\rm coul}(R)$ is consistent with a linear increase, $\sigma_{\rm coul} > 0$, at temperatures $T$ above the phase transition, $T > T_c$, where the physical string tension vanishes, $\sigma = 0$ \cite{Greensite:2004a}.  Investigation of the temperature dependence of $\sigma_{\rm coul}$ revealed that at high $T$ in the deconfined phase, the Coulomb string tension {\em increases} with $T$, consistent with a magnetic mass \cite{Nakagawa:2006},
\beq
\label{magmass}
\sigma^{1/2}_{\rm coul}(T) \sim c \ g^2(T) \ T.
\eeq	
Thus, from the numerical evidence, what has been called the Gribov confinement  scenario works as well in the deconfined phase of QCD as in the confined phase.

	Although this came as a surprise, it was a surprise that could have been predicted.  Recall that the temperature~$T$ determines the extent, $\beta = T^{-1}$, of Euclidean space-time in the time direction.  On the other hand gauge fixing to the minimal Coulomb gauge, defined in~(\ref{lambda}) below, is done at each $t$, so the Coulomb gauge condition (\ref{gaugecond}) holds at each Euclidean time~$t$ and the cut-off at the Gribov horizon applies to 3-dimensional configurations ${\bf A}({\bf x}, t)$ at each time, $t$, regardless of the extent $\beta$ in the time direction.  Thus the arguments which lead to suppression at small~$\bf k$ of the space components $D_{A_i A_j}({\bf k}, k_0)$ of the gluon propagator and enhancement at small~$\bf k$ of the time component $D_{A_0, A_0}({\bf k}, k_0)$ hold at all temperatures~$T$, including in the deconfined phase.  
		
	It thus appears, from both numerical and analytic evidence, that what  originated as a scenario for confinement provides a more general framework for QCD that holds both in the confined and deconfined phases.  Although this is surprising at first, there is no paradox because the bound (\ref{boundV}) provides a necessary condition for confinement, but not a sufficient condition.  
	
	(To see that the converse may be false --- namely that the color-Coulomb energy may diverge even though the quark-pair energy is finite --- suppose that the ground-state quark-pair energy is confining when the quarks are in the fundamental representation, $V_{\rm fund}(R) \sim R$.  In this case according to the bound (\ref{boundV}), $-V_{\rm coul}(R) \to \infty$.  But then the color-Coulomb quark-pair energy $-C_r  V_{\rm coul}(R)$ is confining in every representation $r$ with $C_r \neq 0$.  This includes the adjoint representation which however is not confining.)

\subsection{Present results}

	In a first attempt to account for the influence of the cut-off at the Gribov horizon on the equation of state of the gluon plasma, a simple Planck formula was taken as Ansatz (\cite{Zwanziger:2004b}), with a dispersion relation given by the Gribov energy $E = ({\bf k}^2 + {m ^4 \over {\bf k}^2 })^{1/2}$, where the Gribov mass $m$ was assumed to be temperature-independent.  In the present work we develop a systematic calculational scheme in which the cut-off at the Gribov horizon is effected within the framework of a local, BRST-invariant, renormalizable quantum field theory, and a renormalizable gap equation.  The equation of state is no longer given by a simple Planck formula, and the Gribov mass becomes a calculable quantity that is temperature-dependent, $m = m(T)$.  We calculate the leading correction to the Stefan-Boltzmann equation of state at high temperature due to the cut-off at the Gribov horizon.    
	
	In 1980 Lind\'{e} showed that standard and resummed finite-temperature perturbation theory suffer from infrared divergences~\cite{Linde:1980}.  Since then no solution has been found, although the infrared divergences may be avoided by introducing a magnetic mass for the gluon in an ad hoc manner.  Some time ago it was suggested that these divergences arise because standard perturbation theory neglects the suppression of infrared gluons required by the proximity of the Gribov horizon in infrared directions~\cite{Zahed:1999}.  No infrared divergence appears in the present calculation because the tree-level propagator~(\ref{transprop}) of would-be physical gluons is strongly suppressed in the infrared.  The correction obtained here is of order~$g^6$, which is precisely the order at which the Lind\'{e} divergence would otherwise occur.  
	
	We show by integrating out auxiliary fields that the local, renormalizable formulation presented here is equivalent to the non-local action and gap equation that had been obtained previously from the standard Coulomb-gauge action by imposing a cut-off at the Gribov horizon~\cite{Zwanziger:1997}.  Determination of the number of independent renormalization constants (there are two) will be presented elsewhere \cite{Zwanziger:2007b}.

\subsection{Other approaches}		

	Equal-time correlators in Coulomb gauge may be calculated using the Schwinger-Dyson equations of the Hamiltonian formulation \cite{Szczepaniak:2006, Szczepaniak:2001rg, Reinhardt:2004a, Reinhardt:2004b, Reinhardt:2006}.  The gluon energy obtained by variational calculation accords at both high and low momentum with the Gribov energy (\ref{poles}), provided that the Faddeev-Popov determinant is properly accounted for \cite{Reinhardt:2004a, Reinhardt:2004b}.  Indeed the Faddev-Popov determinant dominates the low-energy dynamics.  A systematic study of the infrared limits of propagators and vertices has been reported recently in \cite{Reinhardt:2006}, which incorporates earlier results \cite{Zwanziger:2001kw, Lerche:2002ep}.   The Schwinger-Dyson equations in Coulomb and Landau gauge should be quite reliable in the infrared limit because the weight in the infrared limit is precisely given by the Faddeev-Popov determinant, $\det(-D_i \p_i)$ with negligible contribution from the gluon wave-functional or action \cite{Zwanziger:2003de}, a phenomenon known as ghost dominance \cite{vonSmekal:1997is, vonSmekal:1997isa}.  This is used to advantage with the horizon condition and non-renormalization of ghost-gluon vertices and, as a result, the infrared limit of the Schwinger-Dyson equations in Coulomb and Landau gauges decouples from finite-momentum correlators.  Moreover errors from truncation of 3-vertices have been controlled \cite{Maas:2005, Reinhardt:2006, Alkofer:2004it}.  By contrast the semi-perturbative (or semi-nonperturbative) method described here may not give accurate results in one-loop approximation for infrared quantities.  Nevertheless both methods have their advantages.  In particular the present approach also gives the correlators in Coulomb gauge at unequal times and finite temperature, and does not rely on a variational Ansatz for the wave-functional that would be needed to go beyond the infrared limit.

	The instantaneous color-Coulomb potential has recently been used in Dyson-Schwinger and Bethe-Salpeter equations to find solutions for pseudoscalar and vector mesons \cite{Alkofer:2006a}.  Cancellation of the energy divergences of Coulomb gauge has been demonstrated at the two-loop level \cite{Taylor:2006}, and renormalization and cancellation of energy divergences to all orders in perturbation theory in Coulomb gauge has been elucidated recently, \cite{Niegawa:2006a, Niegawa:2006b}.

	Equivalence of a cut-off at the Gribov horizon to a modified local action with auxiliary fields was first established in Landau gauge some time ago \cite{Zwanziger:1989}.  It was then shown that the horizon condition renormalizes consistently, and that the ghost propagator in Landau gauge has a ${1 \over (k^2)^2}$ or dipole singularity \cite{Zwanziger:1993}.  The symmetries of the local theory in Landau gauge were exhibited and algebraic renormalizability established  \cite{Maggiore:1994, Dudal:2005}.  More recently, using this local action in Landau gauge, the gap equation was determined to two-loop order, and it was verified at the two-loop level that the ghost propagator has a dipole singularity in the infrared  \cite{Gracey:2005}.  It has also been found to one-loop order that the gluon propagator in Landau gauge vanishes like $k^2$, and that the renormalization-group invariant coupling $\alpha_s(k)$ appropriate to the Landau gauge is finite at $k = 0$~\cite{Gracey:2006}.  Thus the elements of Gribov's scenario in Landau gauge \cite{Gribov:1977wm} have been derived from a local, renormalizable, BRST-invariant action.  The Landau gauge case has been reviewed in \cite{Sobreiro:2005}.

	The Coulomb gauge provides a more straightforward confinement scenario than the Landau gauge (but see~\cite{Alkofer:2006}), and for this reason Gribov and others turned to the Coulomb gauge.  The Coulomb gauge is also well suited to finite-temperature calculations because the heat-bath provides a preferred Lorentz frame, so symmetries of the Coulomb gauge, which breaks manifest Lorentz invariance, are the physical symmetries of the system.  Unitarity is immediate in the non-perturbative Coulomb gauge because of the equivalence to a canonical system whereas it is problematical in the non-perturbative Euclidean Landau gauge.

\subsection{Organization of paper}

	In sect.~II we briefly review previous work on which this article relies, which is scattered in a number of different places.  A local, renormalizable, BRST-invariant action with auxiliary ghosts and sources is introduced in sect.~III.  In sect.~IV the non-perturbative Coulomb gauge is defined by a gap equation that determines a non-zero ``physical value" for sources.  Unitarity and confining properties of the non-perturbative Coulomb gauge are established in sect.~V.  Free propagators are calculated in sect.~VI.  The gap equation is evaluated in one-loop approximation in sect.~VII and solved in sect.~VIII.  The free energy is calculated in sect.~IX, and in sect.~X the leading high-temperature correction to the equation of state from the cut-off at the Gribov horizon is obtained.  Our conclusions are presented in sect.~XI.  The auxiliary fields are integrated out in Appendix~A, and which results in the standard Coulomb gauge with a cut-off at the Gribov horizon.  In Appendix~B the partition function in Coulomb gauge with a cut-off at the Gribov horizon is expressed in canonical hamiltonian form.

\section{How to cut off a functional integral (old stuff)}
	
	In Euclidean quantum field theory the functional integral
\beq
Z = \int_\Lambda dA \ \delta(\p_iA_i) \det[D_i(A) \p_i]
 \exp( - S_{YM}),
\eeq	
must be cut off at the boundary $\p \Lambda$ of the ``fundamental modular region"~$\Lambda$, this being a region on the gauge-fixing surface that intersects each gauge orbit once and only once.  The integrand is the familiar Faddeev-Popov weight Coulomb gauge.

	In Coulomb gauge we may take $\Lambda$ to be the set of configurations $A_i({\bf x})$ each of which is the absolute minimum on its gauge orbit of the Hilbert norm,
\beq
\label{lambda}
\Lambda \equiv \{ \ A_i({\bf x}): 
 ||A|| \leq ||{^g}A|| \ \}.
\eeq
Here $A_i({\bf x})$ is a 3-dimensional configuration, ${^g}A_i = g^{-1}A_i g + g^{-1}\p_ig$ is its gauge transform by a 3-dimensional local gauge transformation $g({\bf x})$, and $||A||^2 \equiv \int d^3x \ |A_i(\bf x)|^2$ is the Hilbert square norm.  This gauge choice is done for each time $t$.  

	At a local or global minimum, the functional
$F_A[g] = ||^gA||^2$ is stationary at $g = 1$, and the second variation of $F_A[g]$ is a positive matrix (all its eigenvalues are positive).  This translates into the Coulomb gauge condition $\p_i A_i = 0$, and the positivity of the 3-dimensional Faddeev-Popov operator,
\beq
- D_i(A) \p_i \geq 0.
\eeq
These two properties characterize a set known as the Gribov region $\Omega$,
\beq
\Omega \equiv \{ \ A:  \p_i A_i = 0, \ {\rm and} \ - D_i(A) \p_i \geq 0 \ \},
\eeq
which is the set of configurations each of which is a local minimum on its gauge orbit of the Hilbert norm.  The set of local minima is larger than the set $\Lambda$ of global minima, $\Lambda \subset \Omega$.

	We do not possess an explicit description of the set~$\Lambda$ of global minima.  However it has been argued from stochastic quantization~\cite{Zwanziger:2003np}, a geometrically correct but unwieldy quantization of gauge theory, that integration over $\Omega$ or over $\Lambda$ gives the same expectation-value for all $n$-point functions with finite $n$.  This is consistent with abelian or center-vortex dominance scenarios, because abelian and center-vortex configurations lie on the common boundary of the Gribov region and the fundamental modular region \cite{Greensite:2004b}.  Either as exact truth, or to provide a model worthy of investigation, we replace the fundamental region~$\Lambda$ by the Gribov region~$\Omega$ in the functional integral,
\beq
Z = \int_\Omega dA \ \delta(\p_iA_i) \det[-D_i(A) \p_i]
 \exp( - S_{YM}).
\eeq

	The boundary $\p\Omega$ of the Gribov region, known as the ``Gribov horizon", is characterized in Euclidean Landau gauge or Coulomb gauge at fixed time (with $D \to D-1)$ by  ``horizon function"~\cite{Zwanziger:1989},
\beqa
\label{horizonfnaa}
 \ G(A) & \equiv  \int d^{D-1}x \  [  &
 g  f^{abc} A_i^b \ (M^{-1})^{ad} \ g f^{dec} A_i^e
 \nonumber  \\
&&  - \ (N^2 - 1)(D -1) \  ],
\eeqa
where $M^{ab} = - D_i^{ab}(A)\p_i$ is the $D-1$ dimensional Faddeev-Popov operator, and $M^{ab}(M^{-1})^{bd} f^{dec} A_i^e = f^{aec} A_i^e$.
The horizon function vanishes on the boundary, $G(A) = 0$ for $A \in \p\Omega$, and is negative inside, $G(A) < 0$ for $A \in \Omega$.  Thus the partition function may be written
\beq
Z = \int dA \ \theta[-G(A)] \ 
 \delta(\p_iA_i) \det[-D_i(A) \p_i]
 \exp( - S_{YM}),
\eeq
where $\theta(-G)$ is the step function,
and a product over all times $t$ is understood.  Configuration space with, for example, a lattice cut-off is a high dimensional space, and classical statistical mechanical arguments apply.  Entropy favors population at the boundary $\p \Omega$ where $G(A) = 0$, and in the last integral we may make the replacement $\theta[-G(A)] \to \delta[G(A)]$.\footnote{As an example of this, the radial probability distribution inside an $N$-dimensional sphere of radius $R$ is given by $r^{N-1}dr$. This gets concentrated on the surface of the sphere, at $r = R$, for $N \to \infty$, thus $\int_0^R dr \ r^{N-1} \to \int dr \ \delta(R-r)$.}  The horizon function $G(A)$ is a bulk quantity, like the hamiltonian $H(A)$ in classical statistical physics.  Just as the microcanonical ensemble is equivalent to the canonical ensemble,
$\delta[H(A) - E] \to \exp[- \beta H(A)]$, where~$\beta$ is determined by the condition $\langle H \rangle = E$, likewise the microcanonical weight $\delta[G(A)]$
is equivalent to the Boltzmann weight $\exp[- \gamma G(A)]$, so partition function is given by \cite{Gribov:1977wm, Zwanziger:1989, Zwanziger:1997}
\beq
\label{Boltzmann}
Z = \int dA  \ 
 \delta(\p_iA_i) \det(-D_i(A) \p_i)
 \exp( - S_{YM} - \gamma S_h),
\eeq  
where $\gamma$ is determined by the horizon condition
\beq
\label{horizoncondbb}
\langle S_h \rangle = 0.
\eeq
Here, as a result of the product over all $t$, 
$S_h$ is the integrated horizon function
\beq
S_h = \int dt \ G[A(t)],
\eeq
where $G(A)$ is given in (\ref{horizonfnaa}).  Because $\langle G[A(t)]\rangle$ is independent of $t$, (\ref{horizoncondbb}) is equivalent to $\langle G[A(t)]\rangle = 0$.  In terms of the free energy, $W = \ln Z$, the horizon condition may be written
\beq
\label{horizoncondbc}
{ \p W \over \p \gamma }  = 0.
\eeq  

	The action $S_h$ is non-local in space, but local in time.  It may be expressed in terms of a local, renormalizable action by integration over auxiliary fields, as has been known for some time in the Landau gauge \cite{Zwanziger:1989}.  Precisely the same argument holds for the Coulomb gauge.  It is written out in Appendix A, where we start with the local, renormalizable action that is given below and integrate out the auxiliary fields to obtain~(\ref{Boltzmann}) and ~(\ref{horizoncondbc}).

\section{Perturbative Coulomb gauge}

\subsection{Faddeev-Popov action}

	For simplicity we shall be interested in pure SU(N) gauge theory at temperature $T$, but there is no obstacle to extending our considerations to include quarks.  It is described by a Euclidean action which, for pure SU(N) gauge theory, is of the form  
\beq
\label{euclidean}
S = S_{YM} + s \Xi
\eeq	
where $S_{YM} = \int d^Dx \ {\cal L}_{YM}$ is the Yang-Mills action, with	
\beq
{\cal L}_{YM} = {1 \over 4} \ F_{\mu\nu}^2, 
\eeq	
\beq
F_{\mu \nu} = \p_\mu A_\nu - \p_\nu A_\mu + g A_\mu \times A_\nu.
\eeq
The $s$-exact term $s\Xi$ is defined below.  Here $g$ is the coupling constant, and we use the notation for the Lie bracket 
$(A \times B)^a \equiv f^{abc} A^b B^c$,
where $f^{abc}$ are the fully anti-symmetric structure constants of the SU(N) group.  The color indices $a, b, c$ are taken in the adjoint representation, $a = 1,..., N^2 -1$.  We shall generally  
suppress the color index, and leave summation over it implicit.  Configurations are periodic in~$x_0$, 
\beq
\label{periodic}
A_\mu(x_i, x_0) = A_\mu(x_i, x_0 + \beta),
\eeq 
with period $\beta = 1/T$, where $T$ is the temperature.  The integral over $x_0$ always extends over one cycle, 
$\int dx_0 \equiv \int_0^\beta dx_0$.  We are in $D$ Euclidean dimensions.   Lower case Latin indices take values, $i = 1, 2, ..., D-1$, while Greek indices take values $\mu = 0, 1, ..., D-1$.

	In the BRST formulation there are, in addition to $A_\mu$, a pair of Faddeev-Popov ghost fields $c$ and $\bar c$ and a Lagrange multiplier field, $b$, on which the BRST operator acts according to
\beqa
\label{BRST1}
s A_\mu & = & D_\mu c; \ \ \ \ \ \ \  \ \ \ \ \  sc = - (g/2) c \times c 
\nonumber  \\
s \bar c & = & ib; \ \ \ \ \ \ \ \  \ \ \ \ \ \ \    s b = 0. 
\eeqa	
It is nil-potent, $s^2 = 0$.  Here $D_\mu$ is the gauge-covariant derivative in the adjoint representation, $D_\mu c \equiv \p_\mu c + g A_\mu \times c$.

	The choice of $\Xi$ is the choice of gauge.  Physics is independent of $\Xi$, provided that it provides a well-defined calculational scheme.  The standard Coulomb gauge is defined by the choice $S_{\rm coul} = \int d^Dx \ {\cal L}_{\rm coul}$
\beqa
\label{coulfix}
{\cal L}_{\rm coul} = s\xi_{\rm coul} & = & s  \p_i \bar c A_i 
\nonumber  \\
& = & i \p_i b A_i - \p_i \bar c D_i c.
\eeqa	
The Lagrange-multiplier field $b$ imposes the Coulomb gauge condition $\p_i A_i = 0$.

\subsection{Auxiliary ghosts}
  
	We first introduce the auxiliary ghosts and related sources in the perturbative Coulomb gauge where their role is trivial and superfluous.  We shall then use them to write a local theory that is equivalent to the standard Coulomb gauge with a cut-off at the Gribov horizon.

	Recall that observables $O$ are in the cohomology of $s$ (namely $s$-invariant operators $sO = 0$, modulo $s$-exact operators, $O \sim O + sX$.)  We may freely introduce additional quartets of auxiliary ghost fields on which $s$ acts trivially, because such fields cannot appear in the cohomology of~$s$,
\beqa
\label{BRST2}
s \phi_B^a & = & \omega_B^a; 
\ \ \ \ \ \ \ \  s \omega_B^a = 0
\nonumber  \\
s \bar\omega_B^a & = & \bar\phi_B^a; 
\ \ \ \ \ \ \ \  s \bar\phi_B^a = 0.
\eeqa	
Here $a$ labels components in the adjoint representation of the global gauge group $a = 1, ..., N^2 -1$, and $B$ is a mute index that is arbitrary for the moment.  The fields $\phi_B^a$ and $\bar\phi_B^a$ are a pair of bose ghosts, while $\omega_B^a$ and $\bar\omega_B^a$ are fermi ghost and anti-ghost.  The BRST method \cite{Baulieu:1985} insures that physics is unchanged if we add to the action an $s$-exact term, $S_{aux} = \int d^Dx \ {\cal L}_{aux} = \int d^Dx \  s\xi_{aux}$ that depends on the auxiliary ghost fields.  The auxiliary lagrangian, adapted to the Coulomb gauge, is taken to be\footnote{More generally we may take the gauge-fixing term to be $s \int d^Dx \ \p_\kappa\bar c \ \alpha_{\kappa\lambda} \ A_\lambda$ and the auxiliary action to be $s \int d^Dx \ \p_\kappa \bar\omega_B^a \ \alpha_{\kappa\lambda} \ (D_\lambda \phi_B)^a$, where $\alpha_{\kappa\lambda}$ is a positive symmetric matrix that may be diagonal \cite{Baulieu:1999}}
\beqa
\label{auxacta}
{\cal L}_{aux} & = &  s \ 
\p_i \bar\omega_B^a (D_i \phi_B)^a
\nonumber  \\
& = & \p_i \bar\phi_B^a (D_i \phi_B)^a
  \\   \nonumber
&& - \p_i \bar\omega_B^a 
[ \ (D_i \omega_B)^a + (g D_i c \times \phi_B)^a \ ].  
\eeqa 
The action and Lagrangian density are now
\beqa
\label{lagdensity}
S & = & \int d^Dx \ {\cal L}
\nonumber  \\
{\cal L} & = & {\cal L}_{YM} + {\cal L}_{\rm coul} + {\cal L}_{aux},
\eeqa
where ${\cal L}_{\rm coul}$ and ${\cal L}_{aux}$ are given in (\ref{coulfix})  and (\ref{auxacta}). 

	This action is renormalizable by power counting.  Symmetries and renormalizability of the analogous action in Landau gauge were established in~\cite{Zwanziger:1993} and~\cite{Maggiore:1994}, and similar considerations hold also in Coulomb gauge~\cite{Zwanziger:2007b}.

\subsection{Auxiliary sources}

	To derive the Slavnov-Taylor identities it is standard to introduce sources $K_\mu$ and $L$ for the BRST-transforms that are non-linear $(K_\mu, sA_\mu) = (K_\mu, D_\mu c)$ and $(L, sc) = (L, (- g/2) c \times c)$.  We also introduce auxiliary sources namely, pairs of bosonic sources, $\bar V_{i B}^a$ and $V_{i B}^a$ and pairs of fermionic sources $N_{i B}^a$ and $\bar N_{i B}^a$~\cite{Zwanziger:1993}.  The extended action with all sources is defined by,	
\beqa
\label{extend}
\Sigma & = & \int d^Dx \ \Lambda
\nonumber  \\
\Lambda & \equiv & {\cal L} + K_\mu sA_\mu 
+ Lsc + \bar V_i D_i \phi 
+ s D_i\bar\omega V_i
\nonumber   \\
&& 
+ D_i \bar\omega N_i
+ \bar N_i sD_i\phi
 + \bar V_i V_i - \bar N_i N_i.
\eeqa
Here the color index $a$ and the mute index $B$ are summed over and suppressed, $\bar V_i D_i \phi \equiv \bar V_{iB} D_i \phi_B$ etc.  We have also added the purely source term $\bar V_i V_i - \bar N_i N_i$, which will allow us to write the horizon condition in a multiplicatively renormalizable form.

\subsection{quantum effective action}

	The partition function is given in terms of the extended action by
\beq
Z(J, K, L, V, \bar V, N, \bar N) = \int d\Phi \ \exp[ - \Sigma + (J, \Phi)].	
\eeq
where $\Phi \equiv (A, c, \bar c, \varphi, \omega, \bar\omega, \bar\varphi)$ represents the set of all fields, and $J$ the corresponding sources.  The free energy is defined by $W(J, K, L, V, \bar V, N, \bar N) \equiv \ln Z$, from which the quantum effective action $\Gamma(\Phi, K, L, V, \bar V, N, \bar N)$ is obtained by Legendre transformation.

	The quantum effective action satisfies the same Slavnov-Taylor identity in Coulomb gauge as in Landau gauge and the proof is the same as in Landau gauge~\cite{Zwanziger:1993},   
\beqa
\label{calS}
  & &  
\int d^Dx \  \Big( {\delta \Gamma \over \delta K }
{\delta \Gamma \over \delta A } 
 + {\delta \Gamma \over \delta L }{\delta \Gamma \over \delta c } 
 + i b {\delta \Gamma \over \delta \bar c }
     \\   \nonumber
&&\ \ \ \ \ \ \ \ \ \ \ \ \ \ \ \ 
+ \omega {\delta \Gamma \over \delta \phi }
+ \bar\phi {\delta \Gamma \over \delta \bar\omega} 
- N {\delta \Gamma \over \delta V}
- \bar V {\delta \Gamma \over \delta \bar N }
\Big) = 0.
\eeqa
We will shortly assign non-zero ``physical" values to the sources $V$ and $\bar V$.  This gives an additional contribution $- \bar V_{\rm ph} {\delta \Gamma \over \delta \bar N }$ to the usual Slavnov-Taylor identity.

\section{Non-perturbative Coulomb gauge}

	In this section we describe a local theory which is shown in Appendix A to be equivalent to the standard Coulomb gauge with a cut-off at the Gribov horizon.

\subsection{Modified local action}
	
	The mute index $B$ on the auxiliary ghosts is now defined to be the pair of indices $B \equiv (c, \mu)$, where~$c$ labels components in the adjoint representation of the global gauge group, $c =  1,..., N^2 -1$, and $\mu$ is a Lorentz index.  With this assignment, the BRST operator acts according to  
\beqa
\label{BRST2}
s \phi_\mu^{ac} & = & \omega_\mu^{ac}; 
\ \ \ \ \ \ \ \  s \omega_\mu^{ac} = 0
\nonumber  \\
s \bar\omega_\mu^{ac} & = & \bar\phi_\mu^{ac}; 
\ \ \ \ \ \ \ \  s \bar\phi_\mu^{ac} = 0,
\eeqa
and the auxiliary lagrangian (\ref{auxacta}) reads
\beqa
\label{auxact}
{\cal L}_{aux} & = &
  \p_i \bar\phi_\mu^{ab} (D_i \phi_\mu)^{ab}
  \\ \nonumber  
&&   - \p_i \bar\omega_\mu^{ab} 
[ \ (D_i \omega_\mu)^{ab} + (g D_i c \times \phi_\mu)^{ab} \ ].  
\eeqa 
[The gauge-covariant derivative and the Lie commutator act on the first color index only, because the second color index is mute, thus 
$(D_i \phi_\mu)^{ac} = \p_i \phi_\mu^{ac} + g (A_i \times \phi_\mu)^{ac}$
where $(A_i \times \phi_\mu)^{ab} \equiv f^{acd}A_i^c \phi^{db}$.]  We further stipulate that the ``physical value" of the sources $V$ and $\bar V$ is not zero but rather at
\beqa
\label{Vphys}
V_{i\mu}^{ab} & = & V_{{\rm ph}, i\mu}^{ab} 
\equiv -  \gamma^{1/2}\delta_{i\mu} \delta^{ab}
\nonumber  \\
  \bar V_{i\mu}^{ab} & = &\bar V_{{\rm ph}, i\mu}^{ab} \equiv  \gamma^{1/2}\delta_{i\mu} \delta^{ab},
\eeqa	
where $\gamma$ is a constant with engineering dimension $m^4$ that will be determined shortly.  All other sources are set to~0, in particular $N_{\rm ph} = \bar N_{\rm ph} = 0$.  For purposes of a semi-perturbative expansion in $g$, described below, we write
\beq
\label{gamma}
\gamma^{1/2} = {m^2 \over (2N)^{1/2}g},
\eeq
and we take $m$ to be of order~$g^0$.  This yields the ``physical" local action
\beq
\label{Sph}
S_{\rm ph} = \Sigma|_{V = V_{\rm ph}, \bar V = \bar V_{\rm ph}, K = L = N = \bar N = 0} = \int d^Dx \ {\cal L}_{\rm ph}
\eeq
where
\beq
\label{Lph}
{\cal L}_{\rm ph} = {\cal L}_{YM} + {\cal L}_{coul} + {\cal L}_{aux} + {\cal L}_m
\eeq
\beqa
\label{Lgamma}
{\cal L}_m & = &  - {m^4 \over 2N g^2} 
(D-1)(N^2 -1)
\\    \nonumber
& & +{m^2 \over (2N)^{1/2} g} \ 
[ \  D_i\phi_i -  s(D_i\bar\omega_i) \ ]^{aa}.   
\eeqa   
and $s(D_i\bar\omega_i) = D_i \bar\phi_i
+  g(D_i c \times \bar\omega_i)$.	
The vacuum free energy is the free energy $W = \ln Z$, with sources set to their physical value,
\beq
W_{\rm ph}(m) = - \Gamma_{\rm ph}(m) \equiv - \Gamma|_{V = V_{\rm ph}, \bar V = \bar V_{\rm ph} }.
\eeq

\subsection{Renormalizable horizon condition}

	The specification of the non-perturbative Coulomb gauge is completed by requiring that $m$ be a stationary point of the vacuum free energy
\beq
\label{horizcond}
{ \p W_{\rm ph} \over \p m } = 
- { \p \Gamma_{\rm ph} \over \p m } = 0.
\eeq	
This is a non-perturbative gap equation that determines~$m$  as a function of $\Lambda_{QCD}$, and eventually also of the temperature $T$.  As shown in Appendix~A, this equation expresses the ``horizon condition" obtained previously~\cite{Zwanziger:1997}.  It is compatible with multiplicative renormalization because $m$ renormalizes multiplicatively, as in Landau gauge~\cite{Zwanziger:1993}.

	From $W_{\rm ph} = \ln Z_{\rm ph}$, and $Z_{\rm ph} = \int d \Phi \ \exp(- S_{\rm ph})$, the horizon condition may be written
\beq
\Big\langle { \p S_{\rm ph} \over \p m } \Big\rangle = 0,
\eeq
where the expectation value is defined by the action $S_{\rm ph}$.
By (\ref{Lgamma}), this reads
\beq
\label{horizoncondb}
\langle  D_i (\phi_i - \bar\phi_i)^{aa} 
 - (gD_i c \times \bar\omega_i)^{aa} \rangle 
=  { 2^{1/2} m^2 \over N^{1/2}g}  (N^2-1)(D-1).
\eeq
The second term vanishes,
\beq
\langle (gD_i c \times \bar\omega_i)^{aa} \rangle = 0,
\eeq
because there is no $\bar c \omega$ term in the action (\ref{lagdensity}) and (\ref{Lph}), and translation invariance implies that the terms $\p_i \phi$ and $\p_i \bar\phi$ do not contribute to (\ref{horizoncondb}).  The horizon condition reads 
\beq 
\label{horizoncond}
 \ \langle f^{abc} A_i^b 
 (\phi - \bar\phi)_i^{ca} \rangle
 = {2^{1/2}m^2 \over N^{1/2} g^2} \ (D-1)(N^2 -1). 
\eeq

\section{Properties of the non-perturbative Coulomb gauge}

\subsection{Unitarity}

	In Appendix~B, the equivalence to a canonical system is established, and with it, unitarity.  Unitarity is already manifest at tree level.  Indeed the only poles in $k_0$ of the tree-level propagators (given below) occur at $k_0^2 + {\bf k}^2 + {m^4\over{\bf k}^2} = 0$, corresponding to the Gribov \cite{Gribov:1977wm} energy $E = ({\bf k}^2 + {m^4\over{\bf k}^2})^{1/2}$ .  This real energy is consistent with a hermitian hamiltonian.    By contrast, the tree-level gluon propagator of the non-perturbative Landau gauge \cite{Zwanziger:1989} has poles at $k^2 +  {m^4 \over k^2} = 0$, where $k^2 = k_0^2 + {\bf k}^2$.  This  corresponds to complex energy $E = ({\bf k}^2 \pm i m^2)^{1/2}$.

\subsection{A confining property}

	Integration over Lagrange multiplier field $b$ produces a functional $\delta$-function that imposes the gauge constraint $\p_i A_i = 0$.  Suppose this integration is done, so $A_i$ is identically transverse.  The auxiliary field $\bar \phi_\mu^{ac}$ is also a Lagrange multiplier, and integration over it yields a functional $\delta$-function that imposes the constraint
\beq
\label{phibarcon}
- \p_i D_i^{ab} \phi_j^{bc} 
- \gamma^{1/2} g f^{cba}A_j^b = 0. 
\eeq
Quantization is done in a periodic box of finite spatial volume $V$.  Integration of the last equation over $V$ at fixed time $t$ kills the first term, which is a spatial divergence, and yields a new constraint satisfied by $A_i$,
\beq
\label{zeromomg}
\int_V d^3x \ A_j^b = 0.
\eeq	
This states that the zero-momentum component of~$A_i$ vanishes.  This is an additional gauge condition on~$A_i$ that expresses a confining property:  a gluon with 3-momentum ${\bf k} = 0$ is forbidden.  In accordance with this condition, the propagator $D_{A_i A_j}({\bf k}, k_0)$, calculated below, vanishes with~${\bf k}$.

\subsection{Automatic cut-off at Gribov horizon}

	In Appendix A it is shown that after integrating out the auxiliary fields the action acquires a term proportional to the ``horizon function"
\beqa
\label{horizonfna}
 \ S_h & \equiv  \int d^Dx \  [  &
 g  f^{abc} A_i^b \ (M^{-1})^{ad} \ g f^{dec} A_i^e
 \nonumber  \\
&&  - \ (N^2 - 1)(D -1) \  ].
\eeqa
If one expands in eigenfunctions of the Faddeev-Popov operator $M(A)$, the functional weight acquires the factor $\exp(- c^2/\lambda_0)$, where $\lambda_0({\bf A})$is the lowest non-trivial eigenvalue of $M(A)$.  By definition of the Gribov region $\Omega$,  $\lambda_0({\bf A})$ vanishes as $\bf A$ approaches the boundary $\p \Omega$ of the Gribov region from within.  This constitutes an ``automatic", non-analytic cut-off of the functional integral at the Gribov horizon.

\section{Free propagators}

	We now develop a semi-pertubative calculational scheme.  For this purpose we treat $m$ as an independent parameter of order $g^0$, and calculate all quantities  perturbatively, including the gap equation, to a given order in $g$.  Then $m = m(g, T)$ is determined by solving the the gap equation (\ref{horizoncond}) non-perturbatively.  
	
	We expand the action (\ref{Lph}) in powers of~$g$,
\beq
S = S_{-2} + S_0 + ... \ .
\eeq	  
The leading term is of order $g^{-2}$,
\beq
\label{minus2}
S_{-2} \equiv - {m^4 \over 2Ng^2} \ (N^2 -1) (D-1)L^3 \beta,
\eeq	
where the spatial quantization volume is $V = L^3$, and the time extent $\beta = T^{-1}$.  Although this term is independent of the fields, it should not be ignored because, when the gap equation is solved for $m = m(T)$, it gives a $T$-dependent contribution to the free energy.  The terms in the action of order $g^0$,
\beq
\label{S0}
S_0 = S_{0,YM} + S_{0,1} + S_{0,2} + S_{0,3},
\eeq	
are quadratic in the fields and determine the ``free" propagators, 
\beq
S_{0,YM} \equiv \int d^Dx \ {1 \over 4} 
(\p_\mu A_\nu^a - \p_\nu A_\mu^a)^2
\eeq
\beq
S_{0,1} \equiv  \int d^Dx \  (i \p_i b^a A_i^a - \p_i \bar c^a \p_i c^a)
\eeq
\beq
S_{0,2} \equiv \int d^Dx \ (\p_i \bar\varphi_\mu^{ab} \p_i \varphi_\mu^{ab} 
- \p_i \bar\omega_\mu^{ab} \p_i \omega_\mu^{ab})
\eeq
\beq
S_{0,3} \equiv \int d^Dx \ 
 { m^2 \over (2N)^{1/2} } \ f^{abc} A_i^b 
 (\varphi_i - \bar\varphi_i)^{ca}.
\eeq
The term $S_{0,3}$ causes a mixing of the zero-order transverse gluon and bose-ghost propagators.  

	To calculate the free propagators, we define the field that mixes with $A_i^b$, 
\beq
\psi_j^b \equiv { i \over (2 N)^{1/2} }   f^{abc}
 (\varphi_j^{ca} - \bar\varphi_j^{ca}).
\eeq 	
The orthogonal component ${ 1 \over (2 N)^{1/2} }   f^{abc}
 (\varphi_j^{ca} + \bar\varphi_j^{ca})$ and other components of $\varphi$ and $\bar\varphi$ do not mix with $A_i$.  Moreover because of the Lagrange-multiplier term $i\p_j b  A_j$, only the 3-dimensionally transverse part $A_j^T$ of $A_j$ contributes to propagators, and it mixes only with the transverse part $\psi_j^T$ of $\psi_j$.  Consequently, the free propagators of the fields $A_j$ and $\psi_j$  are determined by the mixed action
\beqa
\label{mixact}
S_0(A, \psi) & \equiv & \int d^Dx \ \Big( \ (1/2) [ (\dot A_j^T)^2 + (\p_i A_j^T)^2 
+ (\p_i \psi_j^T)^2 ]
\nonumber   \\
&& \ \ \ \ \ \ \ \ \ \ \  \ \ \  -im^2 A_j^T \psi_j^T \ \Big),
\eeqa
which corresponds to the matrix in momentum space,
\[
\left(
\begin{array}{cc}
 {\bf k}^2 + k_0^2 & \ \  -im^2     \\
-im^2  &  \ \  {\bf k}^2
\end{array}
\right),
\]
with determinant
\beq
\label{Det1}
\Delta_1 = ({\bf k}^2 + k_0^2){\bf k}^2 + m^4 .
\eeq
The free propagators are given by
\beqa
\label{glueprop}
D_{A_iA_j}(x-y)  & = &  \int { d^{D-1}k \over (2\pi)^{D-1} } \ 
T\sum_{k_0} \exp[i k \cdot (x- y) ]
\nonumber  \\
&&  \times  P_{ij}({\bf k}) 
{  {\bf k}^2 \over ({\bf k}^2 + k_0^2){\bf k}^2 + m^4 }
\eeqa
\beqa
\label{mixedprop}
D_{A_i \psi_j}(x-y) & = & \int { d^{D-1}k \over (2\pi)^{D-1} } \ 
T\sum_{k_0} \exp[i k \cdot (x- y) ]
\nonumber  \\
&&  \times P_{ij}({\bf k}) 
{  im^2 \over ({\bf k}^2 + k_0^2){\bf k}^2 + m^4 }
\eeqa
\beqa
D_{\psi_i \psi_j}(x-y) \rangle & = & \int { d^{D-1}k \over (2\pi)^{D-1} } \ 
T\sum_{k_0} \exp[i k \cdot (x- y) ]
\nonumber  \\
&&  \times P_{ij}({\bf k}) 
{  {\bf k}^2 + k_0^2 \over ({\bf k}^2 + k_0^2){\bf k}^2 + m^4 }.
\eeqa
Here $k_0 = 2\pi n/\beta$ are the Matsubara frequencies, where $n$ is any integer, $T = 1/\beta$, and $P_{ij}({\bf k}) = \delta_{ij} - \hat k_i \hat k_j$ is the transverse projector.  We have suppressed the trivial color factor $\delta^{bc}$.  The gap equation (\ref{horizoncond}) reads 
\beq 
\label{horizonconda}
 -i \ \langle A_j^c(0) \psi_j^c(0) \rangle
 = {m^2 \over N g^2} \ (D-1)(N^2 -1). 
\eeq

\section{Gap equation in one-loop approximation}

When the left-hand side of the gap equation is evaluated to zeroth order in $g$, using the mixed propagator~(\ref{mixedprop}), it reads
\beq
\label{gaphorizon}
 \int { d^{D-1}k \over (2\pi)^{D-1} } \ 
T\sum_{k_0} 
 { D-2 \over ({\bf k}^2 + k_0^2){\bf k}^2 + m^4 } = {D-1 \over Ng^2 },
\eeq
where we used $P_{ii}({\bf k}) = D-2$.

	To evaluate the sum over Matsubara frequencies,
\beqa
\label{sum}
Q & \equiv & T\sum_{k_0} 
 { 1 \over ({\bf k}^2 + k_0^2){\bf k}^2 + m^4 } 
\nonumber  \\  
& = & {1 \over E^2 {\bf k}^2 \beta } \sum_{n=0, \pm 1...}	
 { 1 \over 1 + (2 \pi n / \beta E)^2 },
\eeqa
where $E \equiv ({\bf k}^2 + { m^4 \over {\bf k}^2 } )^{1/2}$,
we use the identity
\beq
\sinh \theta = \theta \prod_{n=1}^\infty [ 1 + (\theta / n\pi)^2 ]
\eeq	
or
\beq
\ln \sinh \theta = \ln \theta +  \sum_{n=1}^\infty \ln [ 1 + (\theta / n\pi)^2 ].
\eeq	
This gives upon differentiation
\beq
{ \cosh \theta \over \sinh \theta }  = {1 \over \theta } 
+ {2 \over \theta } \sum_{n=1}^\infty {1 \over  1 + ( n \pi / \theta )^2 } ,
\eeq
and we obtain for the sum over Matsubara frequencies
\beqa
\label{sumdone}	
Q & = & { 1 \over 2 {\bf k}^2 E } 
\ { \cosh (\beta E / 2) \over \sinh (\beta E / 2) }
\nonumber  \\
& = & { 1 \over 2 {\bf k}^2 E } \ 
\Big( 1 + { 2 \over \exp(\beta E) - 1 } \Big).
\eeqa
The first term is the value of $Q$ at $T = 0$ and the second is a Planck-type finite-temperature correction.  The gap equation reads
\beq
\label{gapeq}
\mu^{4-D}\int { d^{D-1}k \over  (2 \pi)^{D-1} } \ { (D - 2) \over 2 {\bf k}^2 E }
 \ \Big(1 + { 2 \over \exp(\beta E) -1 }\Big) = { D - 1 \over N g^2 },
\eeq
where we have made the substitution
$g^2 \to g^2 \mu^{4-D}$.   

	The first term on the left hand side, 
\beq
I \equiv \mu^{4-D}\int { d^{D-1}k \over  (2 \pi)^{D-1} } \ { (D - 2) \over 2 {\bf k}^2 E },
\eeq	
converges for $D < 4$.  The integral is readily evaluated by taking $y = |{\bf k}|^4$ as integration variable, with the result
\begin{equation}
I = \Big({m \over \mu}\Big)^{D-4} \ {D-2 \over 2} { 1 \over (4\pi)^{D/2} }
{\Gamma((D-2)/4) \Gamma((4-D)/4) \over \Gamma((D-1)/2) }.
\end{equation}
In terms of $\epsilon \equiv (4-D)/2$, this gives
\begin{equation}
I = {1\over 4 \pi^2}
\Big[ {1 \over \epsilon } - \ln \Big( {m^2\over \mu^2 }\Big) + \ln(4\pi) + {1 \over 2}\Big( \Gamma'(1) + {\Gamma'(1/2) \over \Gamma(1/2) } \Big) +1  \Big],
\end{equation}	
with neglect of terms that vanish with $\epsilon$.  From the identity \cite{JahnkeEmde}, 
\begin{equation}
{\Gamma'(1/2) \over \Gamma(1/2) } = {\Gamma'(1) \over \Gamma(1) } - 2\ln2,
\end{equation}
one obtains
\begin{equation}
I = {1\over 4 \pi^2}
\Big[ {1 \over \epsilon } - \ln\Big( {m^2\over \mu^2 }\Big) + \ln(4\pi) - \gamma - \ln 2 +1  \Big],
\end{equation}
where $\gamma \equiv - \Gamma'(1) = 0.577215... \ $.  Upon making the standard $\overline{MS}$ subtraction, one obtains
\begin{equation}
I \rightarrow {1\over 4 \pi^2}
  \ln\Big( { e \over 2}  { \mu^2 \over  m^2 } \Big)  ,
\end{equation}
After this substitution, the gap equation (\ref{gapeq}) reads
\beq
\label{gapeq2}
 { 1 \over 4  }   \ln \Big( { e \over 2}  { \mu^2 \over  m^2 } \Big) 
+  \int_0^\infty { dx \over u }
\ { 1 \over \exp(m \beta u) -1 }  = { 3 \pi^2 \over N g^2(T) },
\eeq
where $u \equiv (x^2 + {1 \over x^2 } )^{1/2}$.

\section{Solution of one-loop gap equation} 

	To solve the gap equation for $m$, we change unknown from $m$ to $ m^* \equiv m/T$ so it reads
\beq
\label{formgapeq}
f(m^*) \equiv { 1 \over 2  }   \ln \Big(  {1 \over m^* } \Big) 
+  \int_0^\infty { dx \over u }
\ { 1 \over \exp(m^* u) -1 } = y,
\eeq
where
\beq
\label{fofmprime}
y \equiv { 3 \pi ^2 \over N g^2(T) }
  - {1 \over 4} \ln \Big( {e \mu ^2 \beta^2 \over 2} \Big).
\eeq
The function $f(m^*)$ decreases monotonically, $f'(m^*) < 0$, with $f(0) = \infty$ and $f(\infty) = - \infty$, so this equation always has a unique solution 
\beq
m^* = {m \over T} = h(y).
\eeq
Moreover by the inverse function theorem, $h(y)$ is analytic, so in this approximation there is no phase transition.
	
	Asymptotically at high $T$, the running coupling $g(T)$ in the $\overline{MS}$-scheme is given by 
\beq
{1 \over g^2(T)} = {11 \ N \over 24 \ \pi^2} 
\  \ln \Big( { 2 \pi T \over \Lambda_{\overline{MS}} } \Big),  
\eeq
where $\Lambda_{\overline{MS}}$ is a physical QCD mass scale.  We specialize to high temperature $T$ where $g(T)$ is small, so the one-loop gap equation should be a good approximation.  We also suppose that at high temperature $\mu = \mu(T) \approx 2 \pi T$, so	
\beq
\label{yofT}
y(T) \approx { 3 \pi ^2 \over N g^2(T) } - {1 \over 4} \ln \Big( { 2e \pi ^2 } \Big).
\eeq 	
According to~(\ref{formgapeq}) and~(\ref{fofmprime}), large $T$ or small $g$ implies small $m^*$, and for small $m^*$ we have
\beq
\int_0^\infty { dx \over u } \ { 1 \over \exp(m^* u) -1 }
\too { 1 \over m^* } \int_0^\infty dx \  { x^2 \over x^4 + 1 } 
= { \pi \over 2^{3/2} \ m^* },
\eeq
so the gap equation at high $T$ simplifies to
\beq
{ 1 \over 2  }   \ln \Big(  {1 \over m^* } \Big) 
+  { \pi \over 2^{3/2} \ m^* } = y(T).
\eeq
To leading order at high $T$ or small $m^*$, the first term on the left is negligible compared to the second, and we obtain, with $m^* = { m \over T}$ and, with neglect of the second term in (\ref{yofT}),
\beq
\label{mofT}
m(T) = { N   \over 2^{3/2}  \ 3 \  \pi } \ g^2(T) \ T
 \ \ \ \ \ \ \ \ \    T \too \infty.
\eeq
Thus, in the high-temperature limit, $m(T)$ approaches a standard magnetic mass, $m \sim g^2(T) T$.

\section{Free energy}

	To order $g^0$, the free energy $W = \ln Z$ is given by 
\beq
\exp W = \int d \Phi \exp( - S_{-2} - S_0 ),
\eeq
where $S_{-2}$, eq.~(\ref{minus2}), is field-independent and of order $g^{-2}$, while $S_0$, eq.~(\ref{S0}), is quadratic in the fields.  We obtain
\beq
W = W_{-2} + W_0
\eeq
where, for $D -1 = 3$
\beq
W_{-2} = - S_{-2} =  {3m^4 \over 2Ng^2} \ (N^2 -1) L^3 \beta
\eeq
and
\beq
\exp W_0 =  \int d \Phi \exp( - S_0 ).
\eeq

	To evaluate $W_0$ we observe first that the Faddeev-Popov ghost pair contributes a factor $\bf k^2$ which is cancelled by the two factors of $|{\bf k}|^{-1}$ that come from the $A_0$ and $b$ integrations.  Moreover all auxiliary bose and fermi ghost pairs with the same action give contributions to $W^0$ that cancel.  Each single 4-momentum mode of the action $S^0(A, \psi)$, eq.~(\ref{mixact}), contributes $\Delta_1^{-1/2}$ to $\exp W_0$, where, by (\ref{Det1}), $\Delta_1 = ({\bf k}^2 + k_0^2) {\bf k}^2 + m^4$.  Corresponding to the bose ghost $\psi_j^b$ that mixes with $A_j^b$ is an otherwise unpaired fermi ghost mode that contributes $\Delta_2^{1/2}$, where $\Delta_2 = {\bf k}^2$.    The net result is that for each 4-momentum mode of the $A_i^b$ field we obtain a contribution 
$(\Delta_1/\Delta_2)^{-1/2}$, where $\Delta_1/\Delta_2 = k_0^2 + E^2$, and $E = ({\bf k}^2 + { m^4 \over {\bf k}^2 } ) ^{1/2}$.  There is an infinite product over all frequencies $k_0$ for each 3-momentum mode of the $A_i$ field.  For each 3-momentum ${\bf k}$ of the $A_i^b$ field the result of this infinite product is the Planck partition function for a single mode
\beqa
\exp[W_0({\bf k})] & = & \sum_{n=0}^\infty \exp(-nE \beta)
\nonumber  \\
& = & [1 - \exp(- E \beta ) ]^{-1},
\eeqa
\beq
W_0({\bf k}) = - \ln [1 - \exp(- E \beta ) ].
\eeq
The sum over the $N^2 -1$ color modes, the 2 degrees of transverse polarization, and over all 3-momentum modes $\bf k$ yields, with
$\sum_{\bf  k} \too  V \int { d^3k \over ( 2\pi )^3 }$,
\beqa
W_0 & = & - 2 (N^2 -1)V \int { d^3k \over ( 2\pi )^3 }
\ln [1 - \exp(- E \beta ) ]
\nonumber \\
& = & - { (N^2 -1) V \over \pi^2 } \int_0^\infty dk \ k^2 
\ln [1 - \exp(- E \beta ) ]
\nonumber \\
& = & { (N^2 -1)V \beta \over 3 \pi^2 } 
\int_0^\infty { dk \ (k^4 -m^4) \over E \  [\exp(E \beta ) -1] },
\eeqa
where $E = E(k) = (k^2 + { m^4 \over k^2 } )^{1/2}$.  We add the term $W_{-2}$ and obtain to order $g^0$ the free energy per unit volume, $w = W/V$,
\beq
\label{freeenergy}
w = (N^2 -1) \ \beta \ \Big( { 3 \ m^4  \over 2 N g^2 }
+ { 1 \over 3 \pi^2 } 
\int_0^\infty { dk \ (k^4 -m^4) \over E \  [\exp(E \beta ) -1] } \Big).
\eeq
Here $m = m(T)$ is the solution (\ref{mofT}) of the gap equation.

\section{Equation of state at high temperature}

	We now evaluate $w$ in the high-temperature limit, where $g(T)$ is small, and our expansion should be reliable.  From (\ref{mofT}) we obtain	
\beq
\label{freeenergy}
w = (N^2 -1) \ \Big( {  N^3 \ g^6(T)   \over 2^7 \ 3^3 \  \pi^4  }
+ { 1 \over 3 \pi^2 } K(\eta) \Big) \ T^3 ,
\eeq
where
\beq
K(\eta) \equiv \int_0^\infty { dy \ (y^4 - \eta) \over u \  (\exp u  -1) },
\eeq 	
\beq
u \equiv (y^2 + {\eta \over y^2})^{1/2},
\eeq
and
\beq
\eta \equiv \Big({m \over T}\Big)^4 
= \Big( { N \ g^2(T) \over  2^{3/2} \ 3 \ \pi } \Big)^4  
\eeq
is a small parameter.

	The lowest-order expression for $K(\eta)$ at small $\eta$ is obtained by setting $\eta = 0$, 
\beqa
K(0) & = & \int_0^\infty { dy \ y^3 \over  \  \exp y  -1 }
\nonumber  \\
& = &  { \pi^4 \over 15 },
\eeqa
which gives the Stefan-Boltzmann free energy.  To evaluate the leading correction to $K(0)$, we use
\beqa
{ \p K \over \p \eta } & = &  - \int_0^\infty dy \ \Big( 
{ 1 \over u \ (\exp u -1)}
+ { y^4 - \eta \over 2 \ u^3 \ y^2 \ \ (\exp u -1) } 
\nonumber \\ 
&& \ \ \ \ \ \ \ \ \ \ \ \ \ \ \
+  { (y^4 - \eta) \ \exp u \over 2 \ u^2 \ y^2 \ \ (\exp u -1)^2 } \Big). 
\eeqa
We pose $y = \eta^{1/4}x$ and obtain
\beqa
{ \p K \over \p \eta } & = &  - \int_0^\infty dx \ \Big( 
{ 1 \over v \ [\exp(\eta^{1/4}v) -1] }
\nonumber  \\
&& \ \ \ \ \ \ \ \ \ \ \
+ { x^4 - 1 \over 2 \ v^3 \ x^2 \ \ [\exp(\eta^{1/4}v) -1] } 
\nonumber \\ 
&& \ \ \ \ \ \ \ \ \ \ \ 
+  { \eta^{1/4} \ (x^4 - 1) \ \exp (\eta^{1/4}v) 
\over 2 \ v^2 \ x^2 \ \ [\exp(\eta^{1/4}v) -1]^2 } \Big), 
\eeqa
where $v \equiv (x^2 + {1 \over x^2 } )^{1/2}$.  To leading order in $\eta$ this gives
\beqa
{ \p K \over \p \eta } & = &  { - 1 \over \eta^{1/4} }
 \int_0^\infty dx \ \Big( 
{ 1 \over v^2 }
+ { x^4 - 1 \over \ v^4 \ x^2  }  \Big)
\nonumber \\
& = & { - 3 \ \pi \over 4 \cdot 2^{1/2} \cdot \eta^{1/4} }. 
\eeqa
This identity, $K = K(0) + \int_0^\eta {\p K \over \p \eta}$, yields 
\beq
K(\eta) = { \pi^4 \over 15 } - { \pi \ \eta^{3/4} \over 2^{1/2} }. 
\eeq
We obtain for the free energy
\beqa
w & = & (N^2 -1) \  \Big( {  N^3 \ g^6   \over 2^7 \ 3^3 \ \pi^4  }
+ {\pi^2 \over 45 } 
- { N^3 \ g^6   \over 2^5 \ 3^4 \ \pi^4  } \Big) \ T^3,
\nonumber  \\
& = & (N^2 -1) \ \Big( {\pi^2 \over 45 }
- { N^3   \over 10,\!368 \ \pi^4  }  \ g^6(T)  \Big) \ T^3,
\eeqa
which gives the leading correction to the Stefan-Boltzmann limit from the cut off at the Gribov horizon.  

	The equation of state of the gluon plasma follows from the thermodynamic formulas for the energy per unit volume and pressure,
\beq
e = - { \p w \over \p \beta }; \ \ \ \ \ \ \ \ \ \ \ \ \  p = { w \over \beta },
\eeq	 
and entropy per unit volume, $s = {e + p \over T}$.  To calculate the energy density, we use $-\beta {\p g \over \p \beta} = T {\p g \over \p T} = \beta$-function $=  O(g^3)$, which is of higher order.  We thus obtain for the energy density and pressure at high temperature, \beq
e = 3p  = (N^2 -1) \  \Big( \ {\pi^2 \over 15 }
- {  N^3   \over 3,\!456 \ \pi^4  }  \ g^6(T) \Big) \ T^4,
\eeq
with $s = { 4e \over 3T }$.

	This provides the leading correction to the equation of state of the gluon plasma at high $T$ that comes from the cut-off at the Gribov horizon.  Numerically it is small, whereas the correction of order $g^6$ is divergent when calculated perturbatively \cite{Linde:1980}.  To this result must be added the perturbative contributions, including resumations \cite{Kapusta:1989}.  Since standard, resummed perturbation theory diverges at order $g^6$, which is precisely the order of the correction we have found, the result obtained here is consistent with standard perturbative calculations.

\section{Conclusion}

	We have developed a semi-perturbative method of calculation.  It uses the systematic, perturbative expansion of local, renormalizable quantum field theory to calculate all quantities, including the gap equation.  The gap equation is then solved non-perturbatively for the Gribov mass~$m(T)$.  
	 
	We have have applied this method to calculate the leading correction to the Stefan-Boltzmann equation of state of the gluon plasma at high T that comes from the cut-off at the Gribov horizon.  It is of order~$g^6$.  However we have not attempted to calculate the additional perturbative corrections to the Stefan-Boltzmann law to this order~\cite{Kapusta:1989}.  Significantly, $g^6(T)$ is precisely the order at which resummation of ordinary perturbation theory fails due to infrared divergences, as was  first shown by Lind\'e~\cite{Linde:1980}.  Thus our calculation does not contradict the finite results from ordinary perturbation theory and its resummation, but instead gives a finite result precisely where ordinary perturbation theory breaks down.

\smallskip
{\bf Acknowledgements}\\
I recall with pleasure stimulating conversations about this work with Reinhard Alkofer, Laurent Baulieu, David Dudal, Andrei Gruzinov, Klaus Lichtenegger, Atsushi Nakamura, Robert Pisarski, Alexander Rutenburg, Martin Schaden, and Silvio Sorella.  I am grateful to the organizers, Attilio Cucchieri, Tereza Mendes, and Silvio Sorella, of the meeting Infrared QCD in Rio, Brazil, June 5 - 9, 2006, where part of this work was done.

\appendix

\section{Action in terms of horizon function}

	In this Appendix we integrate out auxiliary fields to express the action in terms of the non-local ``horizon function'' that effects the cut-off at the Gribov horizon.  It is non-local in space, but local in time. 
	
	We start with the action (\ref{Lph}).  It contains cross terms that involve $\bar\omega$ but none with $\omega$.  These cross terms are cancelled by an appropriate shift of $\omega$.  We then integrate out the quartet $\omega_0^{ab}, \bar\omega_0^{ab},  \varphi_0^{ab}, \bar\varphi_0^{ab}$ that appears in the action in the expression 
\beq
I_0 = \int d^Dx \ ( \bar\varphi_0^{ac} M^{ab} \varphi_0^{bc} - \bar\omega_0^{ac} M^{ab} \omega_0^{bc}), 
\eeq  
where $M^{ab} = - \p_i D_i^{ab}$ is the Faddeev-Popov operator.  This gives
\beq
\int d \omega_0 d\bar\omega_0 d\varphi_0 d\bar\varphi_0 \exp(-I_0) = 1,
\eeq  
because from the fermi ghost pairs $\omega_0^{ab}, \bar\omega_0^{ab}$, we get $(\det M)^{N^2-1}$ and from the bose ghost pairs  $\varphi_0^{ab}, \bar\varphi_0^{ab}$  we get  $1/(\det M)^{N^2-1}$, which cancel.  (This shows that the auxiliary quartet $\varphi_0, \omega_0, \bar\omega_0, \bar\varphi_0$ plays no dynamical role but it was included because it may be useful to control Lorentz invariance.)  
      
	We now integrate over the remaining auxiliary fermi ghost pairs and obtain
\beq
\int d\omega_j^{ab} d\bar\omega_j^{ab} \ 
\exp(\bar\omega_j^{ac}, M^{ab} \omega_j^{bc}) = 
(\det M)^{(N^2-1)(D-1)}.
\eeq	
The remaining auxiliary bose ghost pairs appear only in the action
\beq
S_{\bar\varphi, \varphi} = \int d^Dx \ 
[ \bar\varphi_j^{ac} M^{ab} \varphi_j^{bc} 
+ \gamma^{1/2} g f^{abc} A_j^b (\varphi_j^{ca} - \bar\varphi_j^{ca}) ],
\eeq	
where we have integrated by parts.  The integral over these ghosts is done by completing the square with the result
\beqa
& \int d\varphi_j^{ab} d\bar\varphi_j^{ab}  & 
\exp( - S_{\bar\varphi, \varphi}) 
 \\  \nonumber
&& = (\det M)^{-(N^2-1)(D-1)} \ \exp( - \gamma  \ S_{h1}), 
\eeqa	
where 
\beq
S_{h1} \equiv \int d^Dx  
 \ gf^{abc} A_i^b \ (M^{-1})^{ad} \ gf^{dec} A_i^e.
\eeq
The powers of $\det M$ from the fermi- and bose-ghost integrations cancel, so the net result from integrating over all the auxiliary ghost pairs is simply $\exp(- \gamma  \ S_{h1})$.

	The only remaining dependence on $\gamma$ occurs in the constant term in ${\cal L}_m$, CORRECT:  	eq.\ (\ref{Lph}).  We combine $S_{h1}$ with the constant term and obtain the non-local horizon function
\beqa
\label{horizonfn}
 \ S_h & \equiv  \int d^Dx \  [  &
 g  f^{abc} A_i^b \ (M^{-1})^{ad} \ g f^{dec} A_i^e
 \nonumber  \\
&&  - \ (N^2 - 1)(D -1) \  ].
\eeqa
The integal over the Faddeev-Popov ghosts, $c$ and $\bar c$, gives $\det M$ as usual, and the integral over the Nakanishi-Lautrup field gives Coulomb gauge condition, $\delta(\p_i A_i)$.  
The partition function is now coincides with (\ref{Boltzmann}).  Moreover, the condition~(\ref{horizcond}) that $m$, or $\gamma$, be a stationary point of the free energy $W = \ln Z$  now reads
\beq
\label{horizcondb}
{ \p W \over \p \gamma } = - \langle S_h \rangle =  0,
\eeq
which coincides with (\ref{horizoncondbb}) and (\ref{horizoncondbc}).
This establishes that the theory described by local, renormalizable Lagrangian~(\ref{Lph}) supplemented by the gap equation~(\ref{horizcond}) is equivalent to the standard Coulomb gauge with a cut-off at the Gribov horizon, as asserted in sect.~II.

\section{Canonical Coulomb-gauge hamiltonian}	
	
	In this Appendix we express the action in terms of the 3-dimensionally transverse field $A_i^T$ and its canonical momentum $\pi_i^T$.  
		
	Although $S_h$ is non-local in space, it is local in time, because the 3-dimensional Faddeev-Popov operator $M$, and thus also its inverse $M^{-1}$, is local in time.  Moreover $S_h$ depends only on $A_i$ but not on its time derivative $\dot A_i$.  The latter appears only in $S_{YM} = \int d^Dx \ {1 \over 4} F_{\mu \nu}^2$ where $F_{\mu \nu}^2 = 2 F_{0i}^2 + F_{ij}^2$ and $F_{0i} = \dot A_i - D_i A_0$.  This allows us to introduce canonical momenta, which are the color-electric fields $\pi_i^a$, by the Gaussian identity
\beqa
\exp( - S_{YM}) & = \int d \pi_i & \exp \Big( - \int d^D x \ 
[ \ {1\over 2} \pi_i^2 
\\  \nonumber   
&& + i \pi_i (\dot A_i - D_i A_0)+ {1 \over 4} F_{ij}^2 \ ] \  \Big),
\eeqa
Integration over $A_0$ imposes Gauss's law in the form $\delta(D_i \pi_i)$, which gives
\beqa
Z && \!  \!  \!  = 
 \int dA_i d \pi_i  \ \det M \ \delta(D_i \pi_i) \ \delta (\p_i A_i) 
\\   \nonumber 
&&  \!  \!  \!  \times \exp\Big( - \int d^D x \ 
[ \ {1\over 2} \pi_i^2 
+ i \pi_i \dot A_i+ {1 \over 4} F_{ij}^2 \ ]    - \gamma \ S_h \  \Big),
\eeqa
We now separate transverse and longitudinal parts, $A_i = A_i^T - \p_i \sigma$ and $\pi_i = \pi_i^T - \p_i U$, where  $\p_i A_i^T = \p_i \pi_i^T = 0$, and  $U$ is the color-Coulomb field, and obtain
\beqa
 \int dA_i d \pi_i \   \det M \ \delta(D_i \pi_i) \ \delta (\p_i A_i) 
 \nonumber  \\ 
 = \int dA^T d \pi^T dU  \  \det M \ \delta(D_i \pi_i),
 \eeqa  
where we have used $\int d\sigma \ \delta(\p_i^2 \sigma) = $ const.  To solve Gauss's law we write 
\beqa
D_i \pi_i & = &  D_i(A^T) ( \pi_i^T - \p_i U)
\nonumber \\
& = & g A_i^T \times \pi_i^T + M(A^T)U,
\eeqa
where $M(A^T) = - D_i(A^T) \p_i$ is the 3-dimensional Faddeev-Popov operator.  The Faddeev-Popov determinant gets absorbed by the identity
\beq
\det M \int dU \ \delta(g A_i^T \times \pi_i^T + MU) = 1.
\eeq
The last $\delta$-function expresses Gauss's law, and fixes the color-Coulomb potential to its physical value,
\beq
U = U_{phys} \equiv M^{-1} \rho,
\eeq
where $\rho \equiv - g A_i^T \times \pi_i^T$ is the color-charge density of the dynamical degrees of freedom.  This gives the partition function in canonical form
\beqa
\label{canonhoriz}
Z & = &  \int dA^T d \pi^T  \ 
 \exp\Big( - \int d^D x \ 
[ \ {1\over 2} \pi_i^2 
+ i \pi_i^T \dot A_i^T
  \nonumber   \\
&& \ \ \ \ \ \ \ \ \ \ \ \ \ \  
+ {1 \over 4} F_{ij}^2(A^T) \ ] - \gamma \ S_h(A^T) \  \Big), 
\eeqa
where $\pi_i \equiv \pi_i^T - \p_i U_{phys}$.  This is a canonical system, and thus unitary, for all real $\gamma$, and stable for $\gamma > 0$.  The canonical Coulomb-gauge hamiltonian of Christ and Lee~\cite{Christ:1980} is modified here by the addition of the term~$\gamma S_h(A^T)$.

	The functional measure $\int dA^T \exp(- \gamma S_h)$,
supplemented by the horizon condition (\ref{horizcondb}) that fixes $\gamma = \gamma_{phys}$, were derived in \cite{Zwanziger:1989} as a representation of the functional integral restricted to the Gribov region,
\beq
\int dA^T \exp(- \gamma_{phys} S_h) 
=  \int_\Omega dA^T.
\eeq
This yields the partition function in terms of the standard Coulomb-gauge first-order action 
\beqa
Z & = &  \int_\Omega dA^T d \pi^T  \ 
 \exp\Big( - \int d^D x \ 
[ \ {1\over 2} \pi_i^2 
+ i \pi_i^T \dot A_i^T
  \nonumber   \\
&& \ \ \ \ \ \ \ \ \ \ \ \ \ \  
+ {1 \over 4} F_{ij}^2(A^T) \ ]  \  \Big), 
\eeqa
where $\pi_i \equiv \pi_i^T - \p_i U_{phys}$, and the integral over $A^T$ is restricted to the Gribov region $\Omega$, defined in sect.~II.  

	Whereas unitarity in Coulomb gauge holds in each order in the perturbative expansion, Lorentz invariance is violated at tree level.  According to the equivalence with the operator formalism, Lorentz invariance is restored only when the non-perturbative gap equation is satisfied, but it is not attempted to verify this by explicit calculation in the present article.

\end{document}